\documentclass[a4paper,fleqn]{cas-sc}

\usepackage[authoryear,longnamesfirst]{natbib}

\def\tsc#1{\csdef{#1}{\textsc{\lowercase{#1}}\xspace}}
\tsc{WGM}
\tsc{QE}
\tsc{EP}
\tsc{PMS}
\tsc{BEC}
\tsc{DE}

\begin{document}
\let\WriteBookmarks\relax
\def\floatpagepagefraction{1}
\def\textpagefraction{.001}
\shorttitle{\textcolor{black}{A Systematic Review of Computational Thinking in Early Ages}}
\shortauthors{Silva et~al.}

\title [mode = title]{\textcolor{black}{A Systematic Review of Computational Thinking in Early Ages}}                      



\author[1]{Edelberto Franco Silva}[type=editor,
                        bioid=1,
                        ]
\cormark[1]
\ead{edelberto@ice.ufjf.br}

\author[2]{Bruno Jose Dembogurski}[type=editor,
                        bioid=2,
                        ]
\ead{brunodembogurski@ic.ufrrj.br}

\author[3]{Gustavo Silva Semaan}[type=editor,
                        bioid=3,
                        ]
\ead{gustavosemaan@id.uff.br}

\cortext[cor1]{Corresponding author}

\begin{abstract}
Nowadays, technology has become dominant in the daily lives of most people around the world. From children to older people, technology is present, helping in the most diverse daily tasks and allowing accessibility. However, many times these people are just end-users, without any incentive to the development of computational thinking (CT). With advances in technologies, the abstraction of coding, programming languages, and the hardware resources involved will become a reality. However, while we have not progressed to this stage, it is necessary to encourage the development of CT teaching from an early age. This work will present state of the art concerning teaching initiatives and tools on programming (e.g., ScratchJr), robotics (e.g., KIBO), and other playful tools (e.g., Happy Maps) for the development of CT in the early ages, specifically filling the gap of CT at the kindergarten level. This survey presents a systematic review of the literature, emphasizing computational and robotic tools used in preschool classes to develop the CT. \textcolor{black}{The systematic review evaluated more than 60 papers from 2010 to December 2020, electing 31 papers and adding three papers from the qualitative stage. The paper's amount was classified in taxonomy to show CT's principal tools and initiates applied to children early. To conclude this survey, an extensive discussion about the terms and authors related to this research area is present.}
\end{abstract}




\maketitle

\sloppy
\section{Introduction}


In this century, people are interested not only in being consumers of technology but also in producing them. Differently to the past decades, today, the children grow up as digital natives. However, to motivate people to not grow only as an end-user, it is necessary to drive them and offer tools to develop computational thinking (CT) as soon as possible. The development of CT is not only essential to create future scientists or engineers, but also it can enhance many cognitive and intellectual skills, allowing people to solve real problems as to ``find the best path from his/her house to the market'', or to ``calculate the trajectory of an object''. 


The CT was credited to Seymour Papert~\citep{papert1980mindstonns} in 1980, but only in 2006 Jeannette M. Wing~\citep{wing2006computational} popularized the term and sparked the international community's interest. Many researches were conducted in the last decade to understand and propose new strategies to develop the CT \citep{bers2017coding,Bers2019,Bers2019b,Palmer2017,Ehsan2018}. However, a little part of these initiatives focuses on early ages, creating a gap to apply STEM education (science, technology, engineering, and math) and develop the CT on the next generations. A overview and updated about the CT in general can be found in Yasar et al. \citep{Yasar2017}. 

\textcolor{black}{
The importance of developing the CT at an early age is discussed by actual papers and books~\citep{bers2020coding, sanford2016computational}. Its primary motivation is to provide a familiarity with CT from an early age, incorporating the benefits to think logically in real-life challenges. However, to exist a motivation to propose and evaluate new curricula and tools from children at an early age, the papers published about this topic are needy. More than this, it is necessary to investigate the state-of-the-art in this field and discuss its characteristics and benefits.
}

\textcolor{black}{
To fill this gap, in this paper, we introduce a systematic review of curriculum initiatives and tools applied to CT from an early age. This work is the first to present the state-of-the-art in this field of knowledge, focusing only on kindergarten and children with 2-5 years old. To conduct our research was considered the decade from 2010 to 2020. Besides presenting the most relevant papers published in the last decade, a growth trend analysis of this research field is introduced as a taxonomy of the main tools used to develop CT at an early age.   
}

A systematic review has clear the emergence of this research topic, showing an increasing number of published papers, a growing  of more than three times in the last five years. The proposal of new curriculum and the development of new tools to support them topics  that needs to be discussed, since a significant number of these are playable using robotic and tangible parts. Another relevant topic is the prevalence of papers in journals concerning conferences, creating the necessity to investigate this topic on symposiums and scientific events. Lastly, the number of papers with case studies and surveys is twice compared to paper without them, reinforcing the need for validation and feedback by education researchers.  


\begin{figure}[htp]
    \centering
    \includegraphics[scale=0.65]{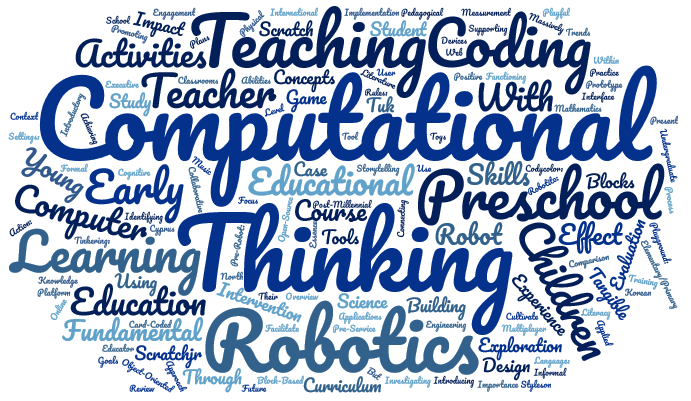}
    \caption{The word cloud shows each word's relative relevance based on the systematic review papers.}
    \label{fig:cloud}
\end{figure}

In Figure \ref{fig:cloud} is shown the most common words across all papers: computational, thinking, programming, and robotics. As aforementioned, we will see in this article, the main initiatives are related to robotics and programming as a way to create playful tools and incentivize the CT at an early age. The word cloud is vital to show the main words and their relevance -- where more bigger words represent more occurrences on the selected papers than small ones. \textcolor{black}{It is relevant to comment about the limitation of this work. In this paper we only consider the gap of CT for the kindergarten level concerning teaching initiatives and tools on programming, robotics, and other playful tools. We do not consider the impact of each initiative applied from an early age to the rest of life (i.e., young and adult life).}

\textcolor{black}{Thus, the main question to be answered by this article is: \textbf{``Which tools and strategies were adopted to encourage the CT development at early ages?''}}


The remainder of this article is structured as follows: section \ref{sec:methods} presents the methodology approach used on this paper, and how the papers were selected; Section \ref{sec:review} describes the leading tools, curriculum initiatives and classify than following a taxonomy proposed by the authors; Section \ref{sec:discussion} conclude this article suggesting a debate and possible future works.


\section{Methods - Systematic Review}
\label{sec:methods}

In this work, a mapping of state of the art was conducted based on a systematic review. This method is responsible for guiding us about collecting the most relevant papers in the last decade, respecting the keywords chosen to answer the question created on the protocol step. The next paragraph explains step by step how we constructed the base of this article. Kitchenham and Charters~\citep{kitchenham2007guidelines} outline three phases to conduct an systematic review: (i) planning; (ii) execution and (iii) summarization/reporting. In the first phase, it is necessary to identify the need for a review and create a review protocol containing the systematic review's important information. The second phase identifies and selects relevant primary studies, performs the data extraction, and synthesizes the extracted data. Finally, in the third phase, the systematic review results are summarized and published to the community. \textcolor{black}{Similar to~\citep{kitchenham2007guidelines}, the PRISMA 2020\footnote{\url{http://www.prisma-statement.org/}}~\citep{moher2011prisma} statement and its checklist and flow diagram}. All steps were conducted using the software StArt~\citep{fabbri2016improvements} that implements all requirements described.

Four indexed libraries were selected, namely: 
\begin{itemize}
    \item Scopus - Elsevier
    \item ACM
    \item IEEE
    \item Web of Science
\end{itemize} 

\begin{figure}[htb]
    \centering
    \includegraphics[scale=0.5]{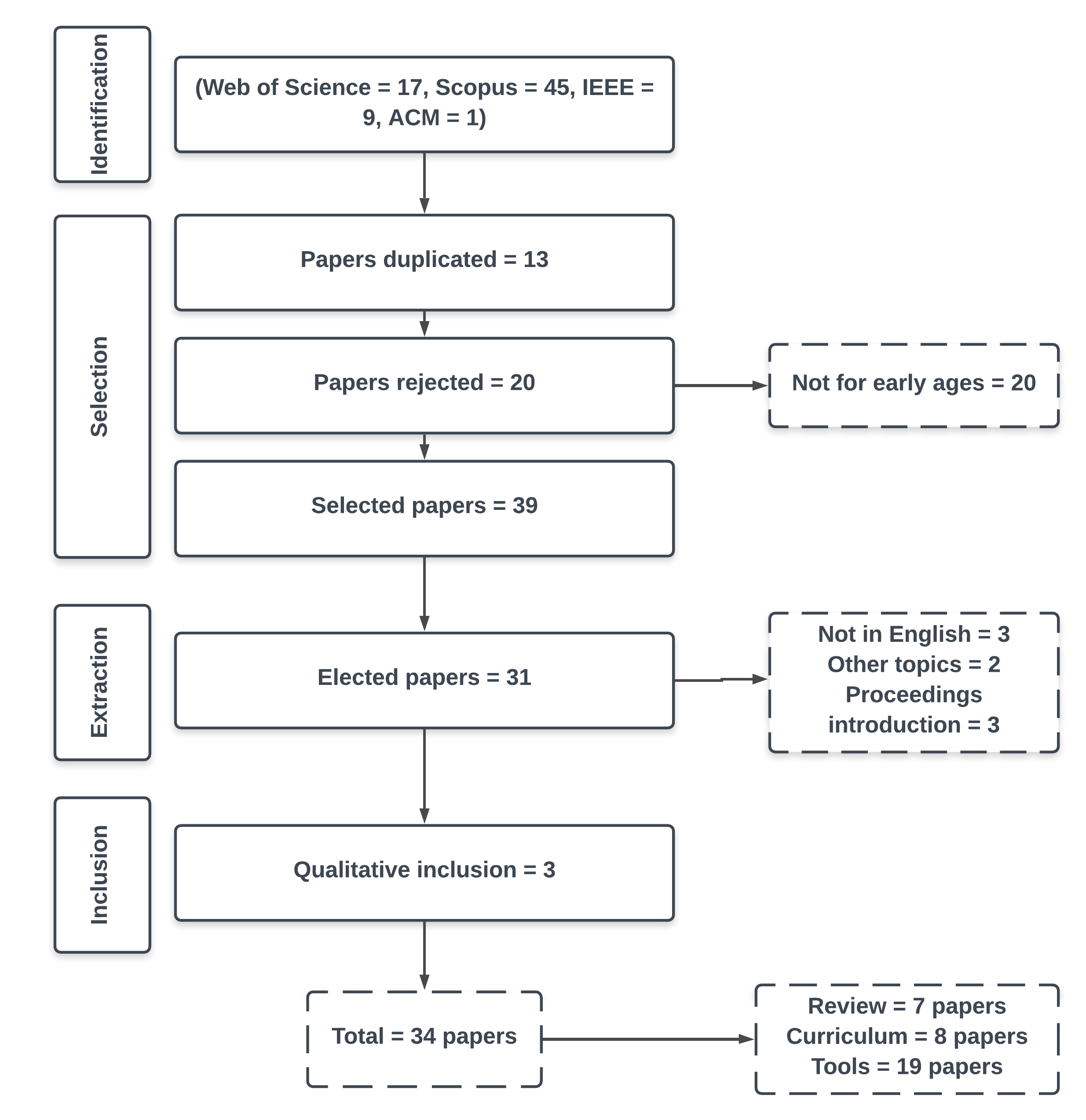}
    \caption{The phases of systematic review following the \textcolor{black}{Kitchenham and Charters~\citep{kitchenham2007guidelines} and PRISMA~\citep{moher2011prisma} guidelines}.}
    \label{fig:review}
\end{figure}

Using the logical keywords combination and restricting the search only to articles and proceedings, written in English and the range of 
\textcolor{black}{
from January 2010 to December 2020
}
, the selection was started.
The search string used on all libraries follows the logic: \texttt{((``computational think*'') AND ((``early age'') OR (``kindergarten'') OR (``preschool'')) AND ((``playful tool'') OR (``programming'') OR (``robotic'')))}, beyond the restrictions already mentioned.

Thus, 72 papers were recovered in the selection step, where 13 papers were excluded as duplicate entries, and 20 rejected them not to be in the scope of ``early ages'' (dealing with children older than five years). With the 39 selected papers, the extraction stage was initiated. In this step, 31 papers were chosen, based on the exclusion of 8 articles because they are outside the standards of full articles in English. Finally, three relevant papers were included in the qualitative inclusion stage, totaling 34 papers. 

\begin{figure}[htb]
    \centering
    \includegraphics[scale=0.5]{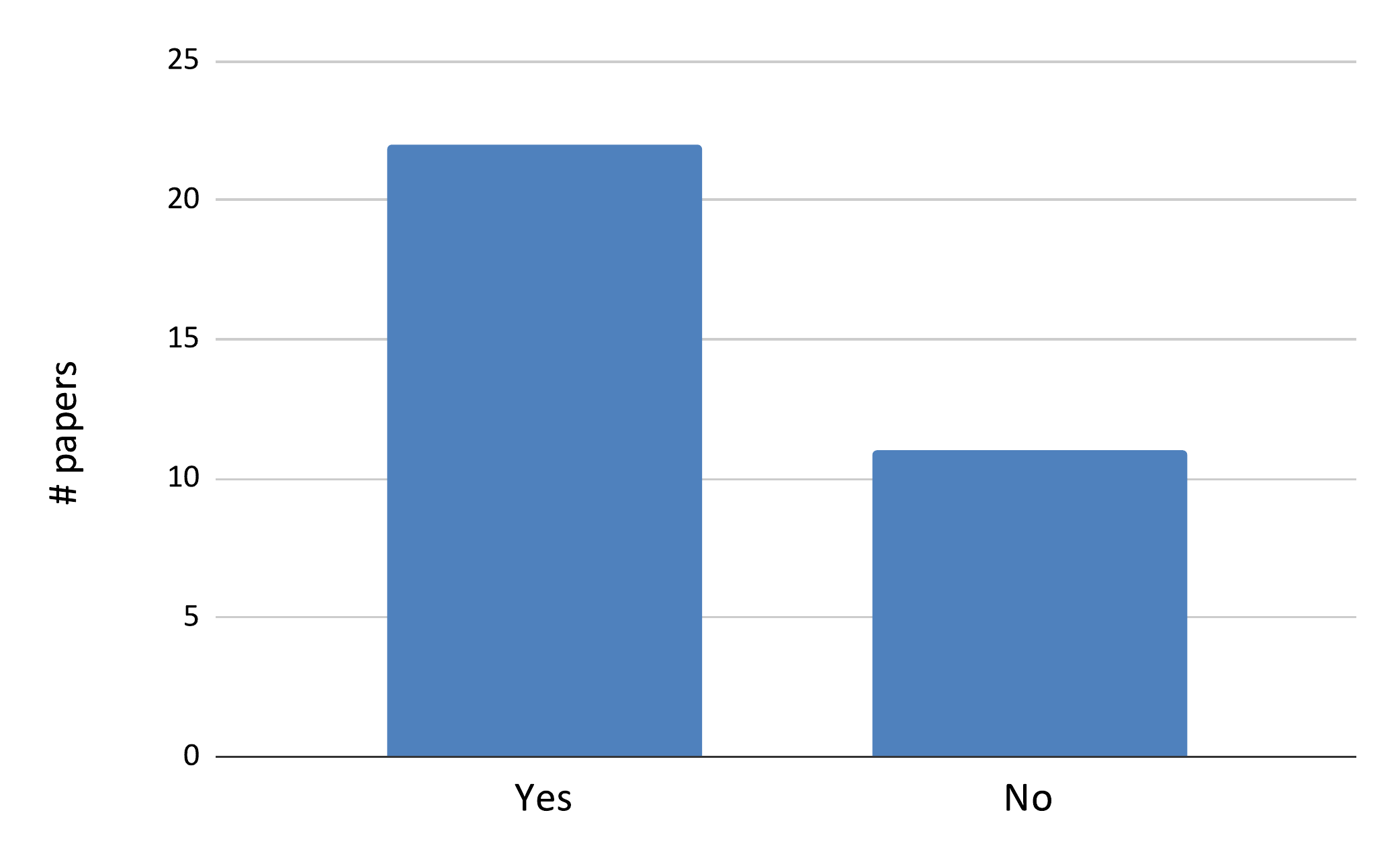}
    \caption{Distribution of paper with (Yes) or without (No) use cases.}
    \label{fig:usecase}
\end{figure}

With this final selection, three were identified having a general review approach to the concept of CT at in early age, seven deal with the proposal for the new curriculum, and 24 modern tools, with or without case studies. It is interesting to note that the case study proves to be of great value in this context and of all works, where 22 articles of the total amount present validation per case study - which represents \~64\% of the papers. Figure \ref{fig:review} shows the steps described in this section, as well as the proportion of paper with (Yes) or without (No) use cases in Figure \ref{fig:usecase}.

\begin{figure}[htb]
    \centering
    \includegraphics[scale=0.5]{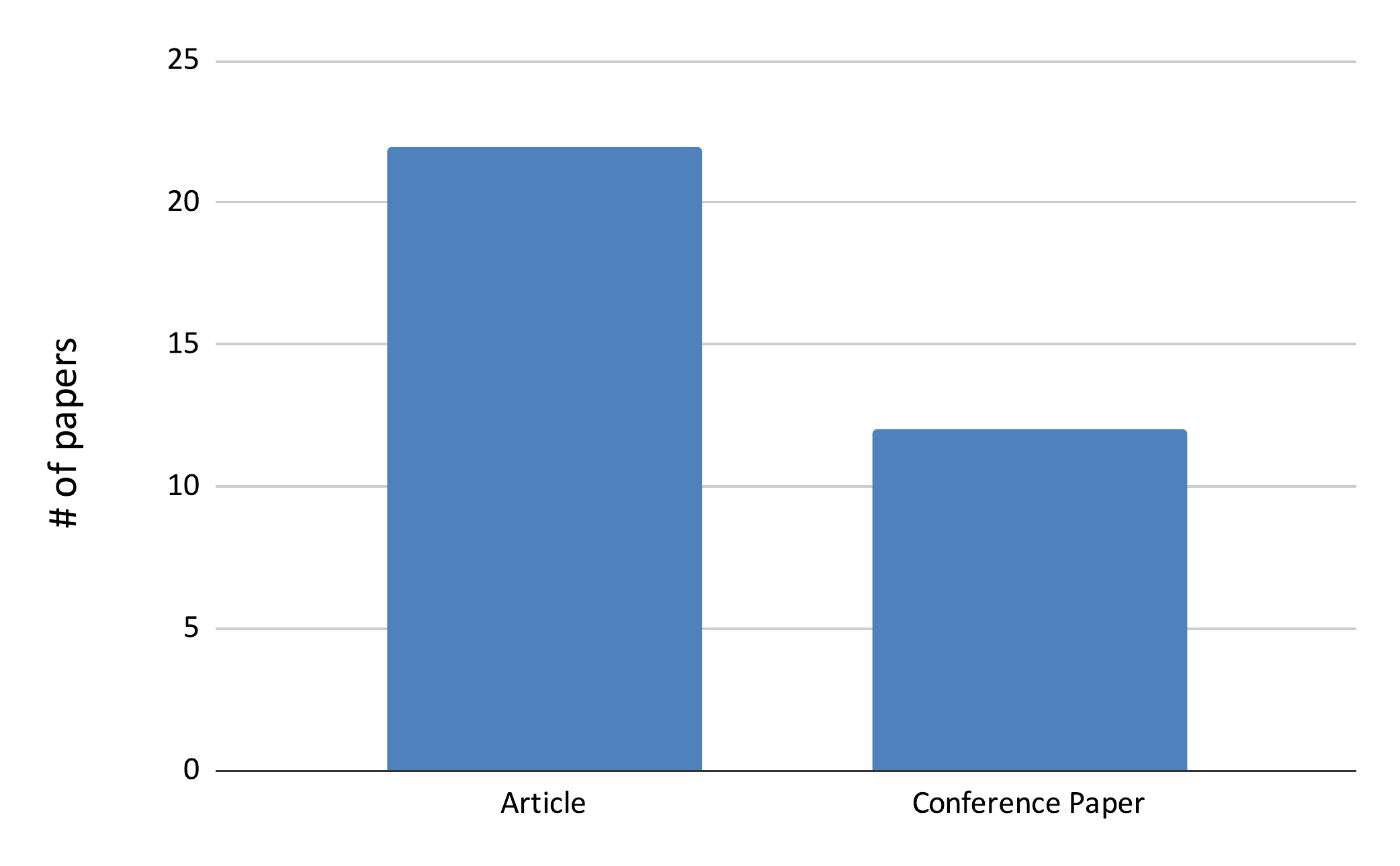}
    \caption{Number of papers published on conference versus journals.}
    \label{fig:count}
\end{figure}

\begin{figure}[htb]
    \centering
    \includegraphics[scale=0.4]{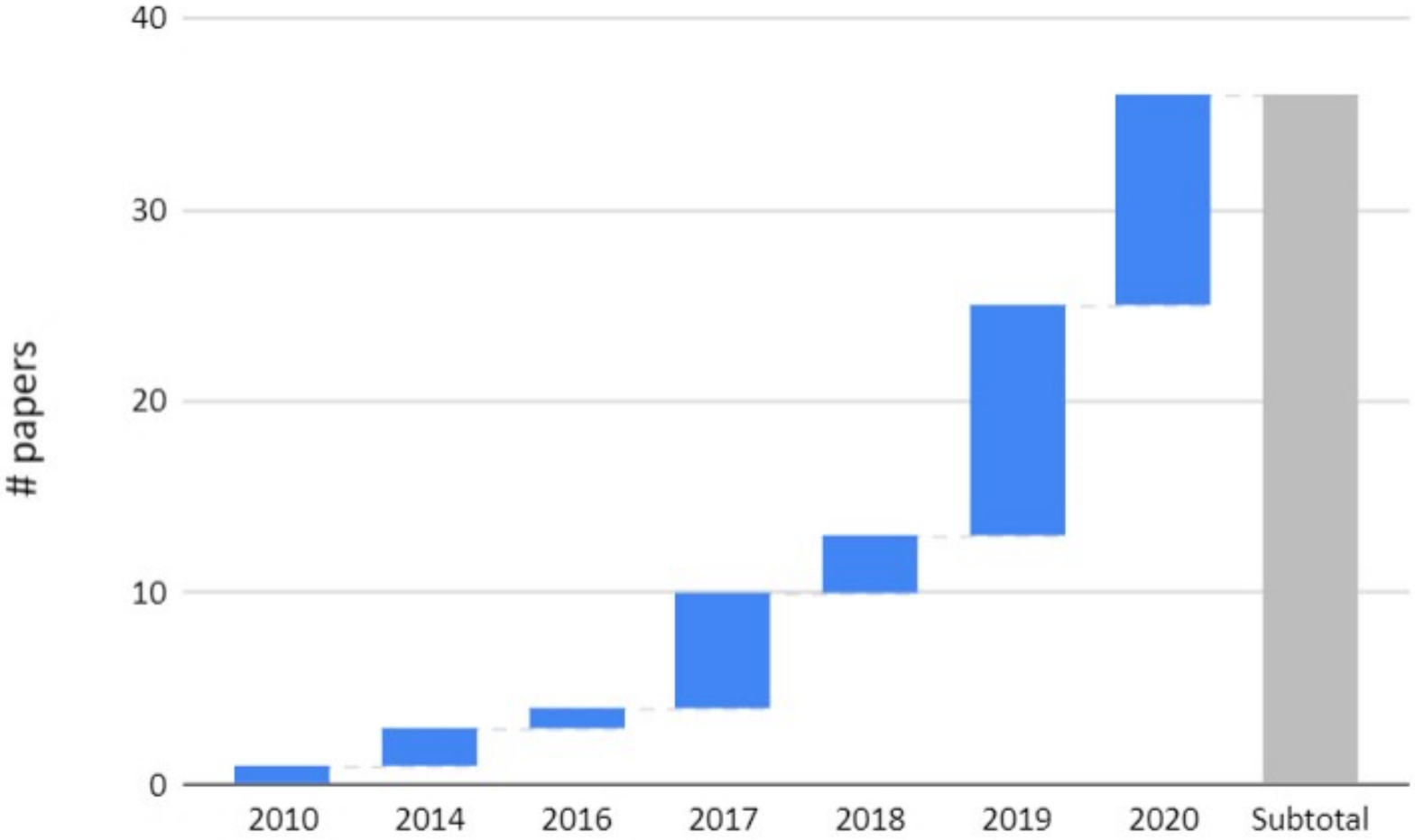}
    \caption{The graph shows the cumulative number of papers per year and its growth trend.}
    \label{fig:trend}
\end{figure}

It is noteworthy that the majority number of papers on journals than conferences, as we can see in Figure \ref{fig:count}. This kind of behavior is not observed in many other research fields. It creates a feeling about the necessity to encourage all people involved in CT to create more forums to discuss the topic, e.g., on symposiums or conferences.

Another relevant preliminary analysis is the relevance of use cases. The majority of the papers present a kind of validation, as through surveys or observations. The distribution of papers with or without use cases can be seen in Figure \ref{fig:usecase}.

As previously mentioned, we will present the main initiatives for the development of CT at an early age. This article is the first research that we know that addresses the topic precisely at an early age. However, we can see that the number of studies and publications has grown in recent years, according to Figure \ref{fig:trend}. The last five years are more than three times bigger than the other years. It is possible to see the growth trend of this research field.

This section introduced how the systematic review was conducted and showed the main characteristics of the papers and their trends. Thus, we can present all papers in the next section.


\section{Results}
\label{sec:review}

The following sections will present the main existing tools, guiding the reader on the selected papers the articles with the proposals and evaluations of the school curriculum, and describing a taxonomy for all of the other papers.

\subsection{Tools}

As part of the initiatives and efforts to develop computational thinking in early ages, tools such as robotics and computer programming initiatives are increasingly being encouraged among researchers and early childhood educators. Robotics and computer programming in this context can support a range of cognitive and social aspects. We will introduce the leading technology solutions that were cited in the selected researches. Basically, in the early ages, there is a great incentive to use block-based programming, a playful way to encourage the child. There is also an incentive to use small robots that memorize simple commands like moving forward, side or back, to create a spatial vision and cognition capacity. There are web-based tools, as we will see, but these are more complex for children at an early age, and there are block solutions that use both the tangible form and the aid of a device such as a tablet or iPad.

\begin{figure}[htb]
    \centering
    \includegraphics[scale=0.5]{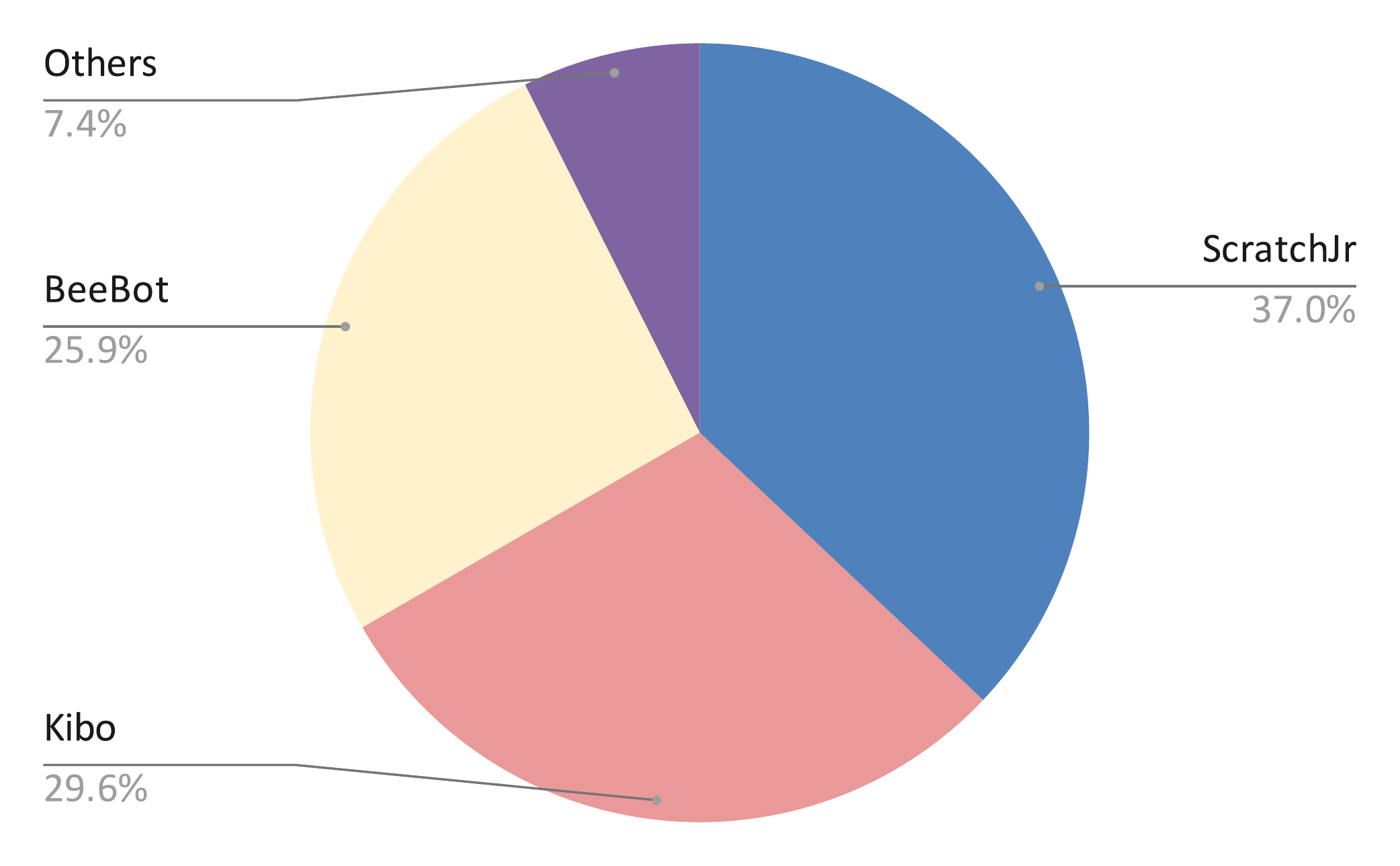}
    \caption{Percentage of main tools occurrence.}
    \label{fig:tools}
\end{figure}

As a way of organizing the presentation and tools' classification, Figure \ref{fig:tools} is presented where it is possible to see BeeBot, Kibo, ScratchJr, and a group of several other tools that appear less frequently in the verified studies. Thus, we will present each one of them below.

\subsubsection{Bee-Bot}

\begin{figure}[htb]
    \centering
    \includegraphics[scale=0.5]{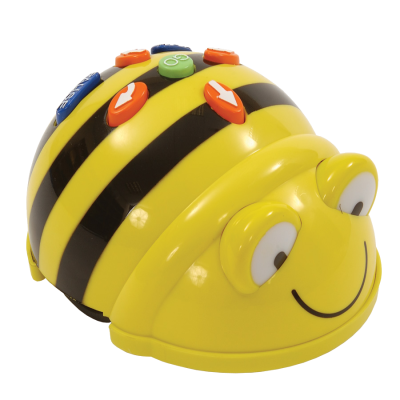}
    \caption{A general view of Bee-Bot with their commands on the top.}
    \label{fig:beebot01}
\end{figure}

The Bee-Bot\footnote{\url{http://bee-bot.us}} is a simple robot based on simple commands focusing on early ages children. This tool is capable of improving the children's skills and develop the CT at an early age, only using a couple of sequences. Their main idea is to enhance the cognition capacity and future view of steps through directional language as a programming sequence: forward, backward, left, and right 90 degree turns. In Figure \ref{fig:beebot01} is possible to see the robot and their commands on top of then. Thus, the sequence is created by the child before to click on the start. After it, the tool enables his/her to run all commands sequentially storage.

\begin{figure}[htp]
    \centering
    \includegraphics[scale=0.4]{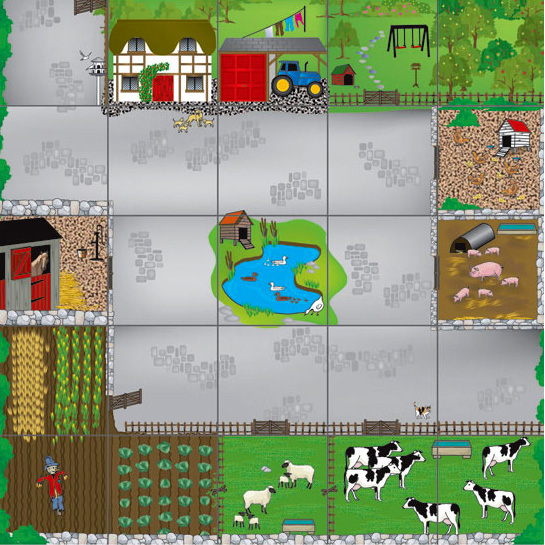}
    \caption{An example of Bee-Bot map.}
    \label{fig:beebot02}
\end{figure}

This robot works as a playful tool once it improves the CT skills through the necessity to memorize the steps and imagine how the robot will run the actions. It is possible to create a puzzle for children using a square paper or a map with obstacles. An example of this kind of problem is shown in Figure \ref{fig:beebot02}.

\subsubsection{KIBO}

\begin{figure}[htb]
    \centering
    \includegraphics[scale=0.4]{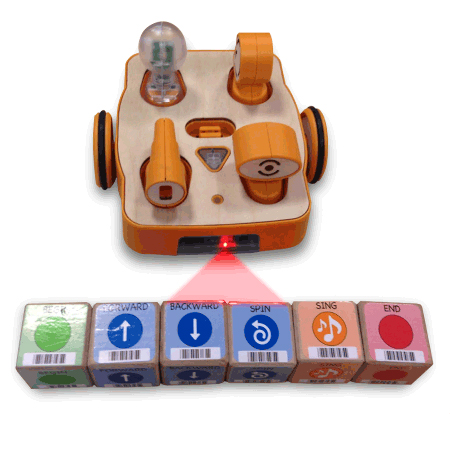}
    \caption{KIBO robot tool with the main blocks reading the code bar by infrared \citep{elkin2016programming}.}
    \label{fig:kibo01}
\end{figure}

The KIBO robotics kit is a tool~\citep{sullivan2015kibo} that propose to engage young children in both building and programming. The KIBO is a kit developed by the DevTech Research Group at Tufts University\footnote{\url{https://sites.tufts.edu/devtech/research/kibo-robot/}} and commercialized by KinderLab Robotics\footnote{\url{https://kinderlabrobotics.com/}}, an enterprise created with a focus do enhance the STEM. Although the KIBO designed for young children with ages from 4 to 7 to learn foundational engineering and programming content, it can be applied to help do develop CT at an early age. Figure \ref{fig:kibo01} is possible to see its main components, the robot, and its blocks with several actions. Once the blocks are read, the robot can run the steps sequentially. It is possible to create another logical programming with the blocks beyond to go forward or change direction; one of these examples is the loop block, mainly used to develop procedures more complex. Melin et al.~\citep{elkin2016programming} present a case use with children with three years old, where the tool is used to validate a proposal to incorporate than on an urban public preschool in Rhode Island, US. 

\subsubsection{ScratchJr}

The ScratchJr\footnote{\url{https://www.ScratchJr.org/}} is a programmable tool web-based free of charge that can be used since the early ages. But this solution is not so simple than KIBO or Bee-Bot needing more help from an adult to understand and manipulate the computer/tablet. Even focusing on children with 5-7 years old, it is common to find many use cases to adopt this tool.

\begin{figure}[htb]
    \centering
    \includegraphics[scale=0.5]{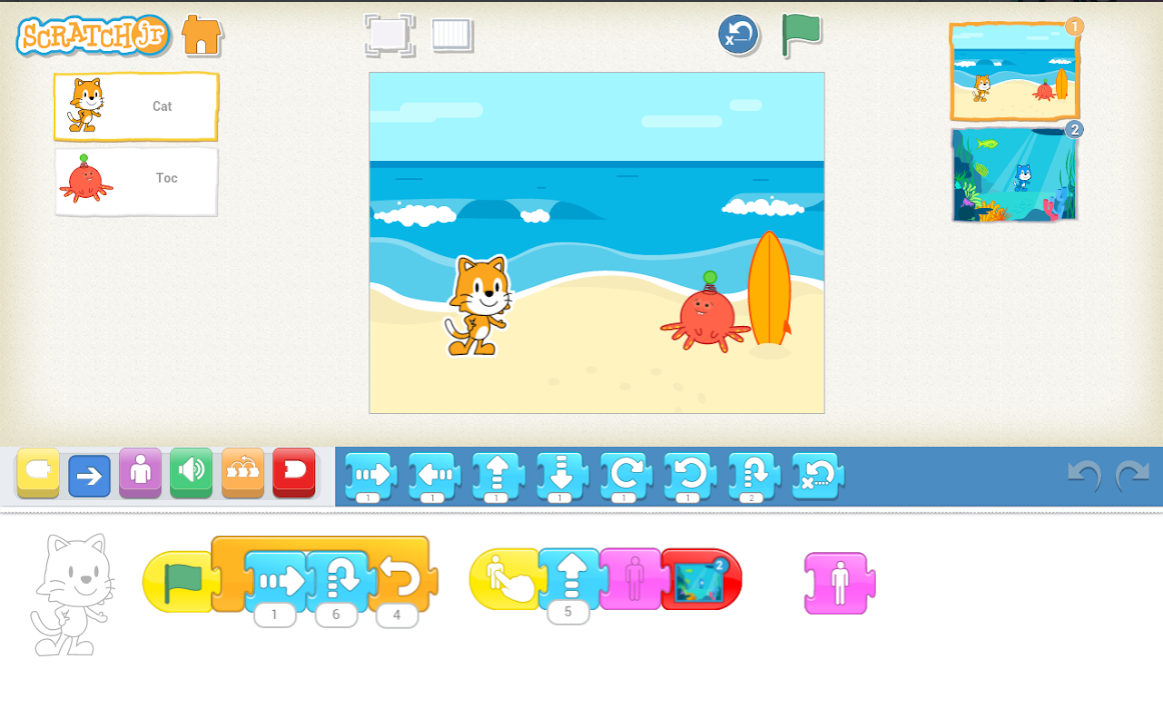}
    \caption{ScratchJr.}
    \label{fig:ScratchJr}
\end{figure}

Its motivation is to improve logic programming from an early age. Like its version for young people, this tool has the purpose of creating funny stories using a sequence of blocks. In Figure \ref{fig:ScratchJr} is possible to see its web interface where a chain of blocks is put on the way to create the story. Although to be more complex, it is possible to create more simple stories with a limited number of blocks, making it possible to be applied at an early age. 

\subsubsection{Others}

\begin{figure}[htb]
    \centering
    \includegraphics[scale=0.4]{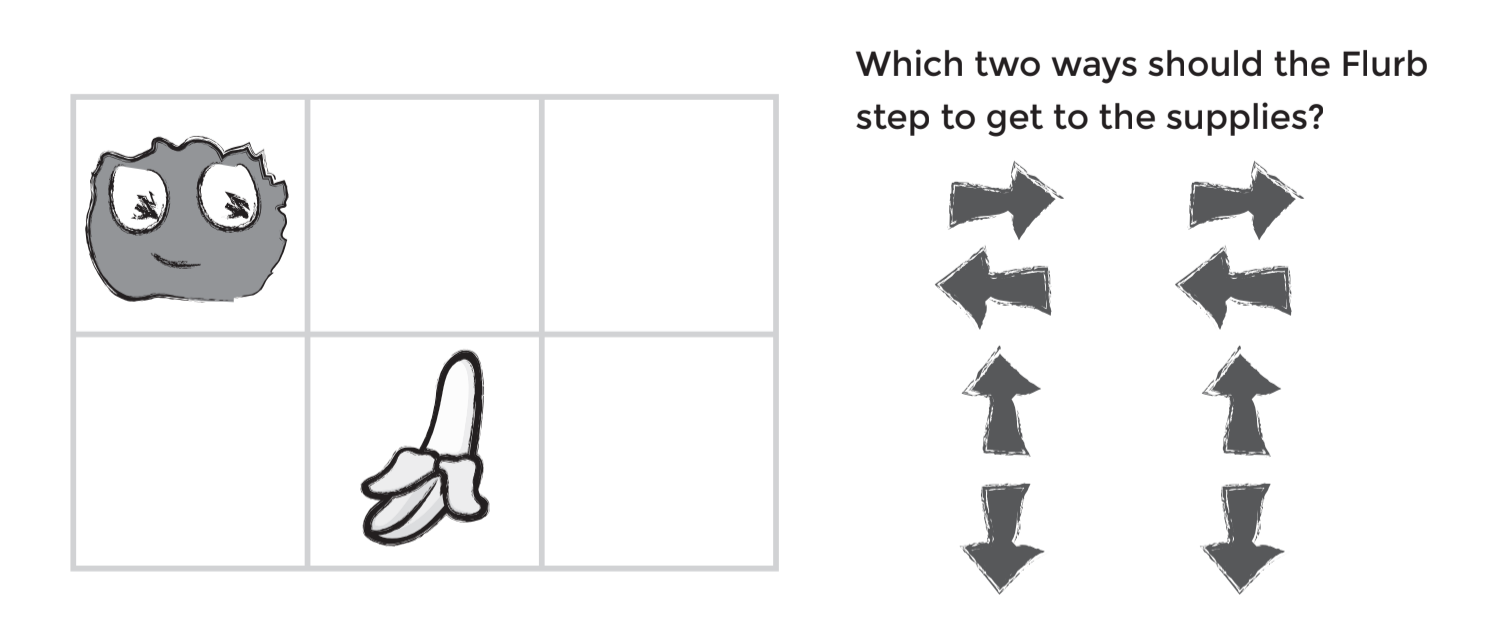}
    \caption{Happy map with a simple example of puzzle to solve a maze - Source: \url{http://code.org}.}
    \label{fig:happy}
\end{figure}

Other tools are cited by many papers, since prototypes of robots to tangible and web-based block-based solutions. Examples of robotic tools are LEGO WeDo and EV3, both are solutions of LEGO company, and their objective is to evaluate the CT for STEM. Despite being on the field of this work, we consider both of them more indicated to be incorporated in the curriculum for children with more than eight years old because of the necessity to manipulate small parts and have the necessary knowledge to manage the computer. Other papers cited the mBot, but it is more like LEGO WeDo. Differently of the tools mentioned, other work proposes a prototype of a playful robotic tool called Robotito. This initiative follows the same KIBO concepts, but today it is only in an initial stage. Another new robot is BlueBot that does the same things as Bee-Bot, but with Bluetooth connection support. Other initiatives as LightBot follow the same of ScratchJr, doing possible to programming with blocks on a web-based application, and can be used for children with five years, for example. More interest to the early ages in this section is the Happy Maps, an unplugged activity. The children can create simple algorithms as a set(s) of instructions to move a character through a maze using a single command. Figure \ref{fig:happy} shows an example of a maze that can be solved by arrows that indicate how the character finds the fruit.

They are a diverse number of tools and prototypes in the literature, and it does not enhance the discussion at this moment. About it, the other tools will be suppressed at this moment. If a tool that was not introduced in this section and appeared ahead, it will be presented in its context.

After knowing about the leading tools in the literature, it is possible to introduce the papers where a curriculum strategy was proposed or validated.  

\subsection{Curriculum}

Many countries have taken initiatives to introduce CT and STEM skills into the classroom, changing the curriculum mainly for people since the K-10 or K-11. It is possible to mention a couple of examples. But in this work, we are focusing since 2 to 8 years old. In the US, the Computer Science For All was created in 2013, to teach students from childhood to high school computer science. In 2014, the European Commission sent a joint letter to the EU Education Ministers, urging them to promote computer programming. In 2012, The Royal Society published a report showing an unsatisfactory result about CT skills development initiatives, proposing a curriculum remodeling. The consequence was the incorporation of the new subject ``Computing'' in primary school (children from 5 to 7 years). France has announced the introduction of an optional programming course in primary schools. Finland and Italy are considering codification initiatives for young people. In Spain, many efforts are being conducted as a non-profit initiative called Programamos, whose fundamental objective is to promote CT's development from an early age through the programming of video games and applications in all school stages. To illustrate it, a set of papers dedicated to propose or validate a curriculum will be present below.


In Bers et al. \citep{Bers2010} was present for the first time the TangibleK, complemented by another paper published in 2014 \citep{Bers2014a}. TangibleK is an education program that uses robotics as a tool to engage children in developing computational thinking and learning about the engineering design process. Supported by the NSF (National Science Foundation), the educational program emerged in 2010 as a perspective for the insertion of computing and CT for children in kindergarten. In \citep{Bers2014a} was conducted a use case considering the five years of learning since the education program' proposal. The program intends to be a guide/framework to assist in developing CT, without being tied to robot technologies or programming languages. The authors highlight the project's results ans its great innovations and help to change the curriculum in schools. Responsible to conduct this research, Prof. Marina Umaschi Bers, from Tufts University\footnote{\url{https://sites.tufts.edu/devtech/}} describe that program as a research involving three dimensions: theoretical contributions, design of new technologies, and empirical work to test and evaluate the theory and the techniques. This research group is the same responsible to create programming languages such as KIBO and ScratchJr, as well as teaching materials and pedagogical strategies for early childhood educators' professional development and community engagement. 

Palmer~\citep{Palmer2017} presents a teaching intervention carried out over four months in two Swedish preschools, where programming was used to facilitate preschoolers’ learning of mathematics, especially in their development of spatial thinking. This study focuses on children between 3 to 4.5 years old, five girls and three boys. Although its small number of participants, the methodology adopted can be considered a dominant feature. The study contains three steps: pretest, intervention, and posttest. The pretest focusing on instruct the child to know about to follow instructions about how to move in a room (i.e., walk three steps in front and turn left); the intervention is composed of four steps for four weeks. The posttest is conducting in rounds after each step, opening an opportunity of each child talks about was learned. 

In Ehsan et al. \citep{Ehsan2018} is present a research as part of the project - supported by NSF - that analyzes the PictureSTEM curriculum\footnote{\url{http://picturestem.org/}}. The authors consider the K-12 STEM to incorporate CT in pre-college education. In this way, the authors look for the ways K-2 children engage in CT in school and out-of-school settings, applying a theoretical framework to find and help CT's development. After conducting the case study, the authors conclude that it provides evidence that children can engage in CT competencies in different problem-solving contexts, including STEM, particularly engineering.  

Munoz-Repiso et al. \citep{GarciaValcarcelMunozRepiso2019} evaluate the repercussion of educational robotics activities on kindergarten students in the acquisition of CT and programming skills. The research design is quasi-experimental, with pre-test and post-test measures, using experimental and control groups. The sample consists of 131 students from the second cycle of early education (between 3 and 6 years old), all from the same Spanish school in Salamanca's city. This initiative is based on the reference program of robotics studies TangibleK.

Recently, Prof. Marina Umaschi Bers, one of the most enthusiastic of CT development at an early age, presents in \citep{Bers2019b} a pedagogical approach for teaching computer science in early childhood. This paper describes an educational strategy for early childhood computer science called ``Coding as Another Language'' (CAL), six coding stages, or learning trajectories that young children go through when exposed to the CAL curriculum. CAL is grounded on the principle that learning to program involves learning how to use a new language (a symbolic system of representation) for communicative and expressive functions. Besides indicating the methodology and how the activities can be incorporated into the curriculum, case studies of young children using either the KIBO robot or the ScratchJr were used to characterize each coding stage and illustrate the instructional practices of CAL curriculum.

In Nam et al. \citep{Nam2019} was examined the effects of a card-coded robotics curriculum and associated activities on kindergarteners’ sequencing and problem-solving skills. Kindergarteners participated in card-coded programming using a robot called TurtleBot\footnote{\url{https://www.turtlebot.com/}}. A card-coded robot curricular intervention was also designed to enhance their planning behaviors using complementary tools. This study examined an 8-week robotic curricular intervention through assessment of 53 participants ranging in age from 5 to 6 in South Korea, while also evaluating sequencing and mathematical problem-solving in both the treatment and comparison groups. It was found that children in the treatment group who engaged in the card-coded robotic curricular intervention performed better on sequencing and problem-solving tests. 

\subsection{Review}

This section is introduced to show some related work about surveys/evaluation using tools or CT curriculum proposals. Considering the scope of this paper, nor have a high number of related work in this area, it is possible to see surveys and studies about tools and CT strategies in pre-schools. As well as the "Curriculum" group, this section is considered an \textit{external box} at taxonomy proposed but keeps its correlation with the research theme.



Umam et al. \citep{Umam2019} reviewed the advantages and disadvantages of commercially available robotics devices. The study shows a critical analysis method to the literature published from 1952 to 2017 in international journals and proceedings. Although only three robotic solutions were investigated, and the discussion is limited to show an overview of the tools.

In Rich et al. \citep{Rich2019} the authors carry out a 20-minute snowball study with groups that teach disciplines related to computing in schools. Around 300 teachers and 60,000 primary school students were interviewed (less than 5 years and up to 14 years, separated into groups 2 years apart). Participants are mainly from the USA and the UK, but also other countries. The most used tools at an early age were: ScratchJr, KIBO, and Unplugged Activities. The authors reach conclusions with their research as: (i) students are capable; (ii) it is ok to fail; (iii) there are a lot of resources to teach computing; (iv) start simple and teach basics; (v) unplugged activities is a good strategy.

Differently, Esteve-Mon et al. \citep{EsteveMon2019} present a study to investigate how to prepare future teachers of Kindergarten and Elementary School in the educational use of CT. Their proposal was conducted on the field of educational robots that includes unplugged, playing, making, and remixing activities. A survey was conducted with 114 Spanish university students of education. The authors divide the strategy into four steps: (i) unplugged, e.g., cut and paste (ii) test of the robots, e.g., Bee-Bot; (iii) activities to ScratchJr individual and group tasks; (iv) and a final mission to resolve a challenge.

\subsection{Taxonomy}

\begin{figure}[htb]
    \centering
    \includegraphics[scale=0.4]{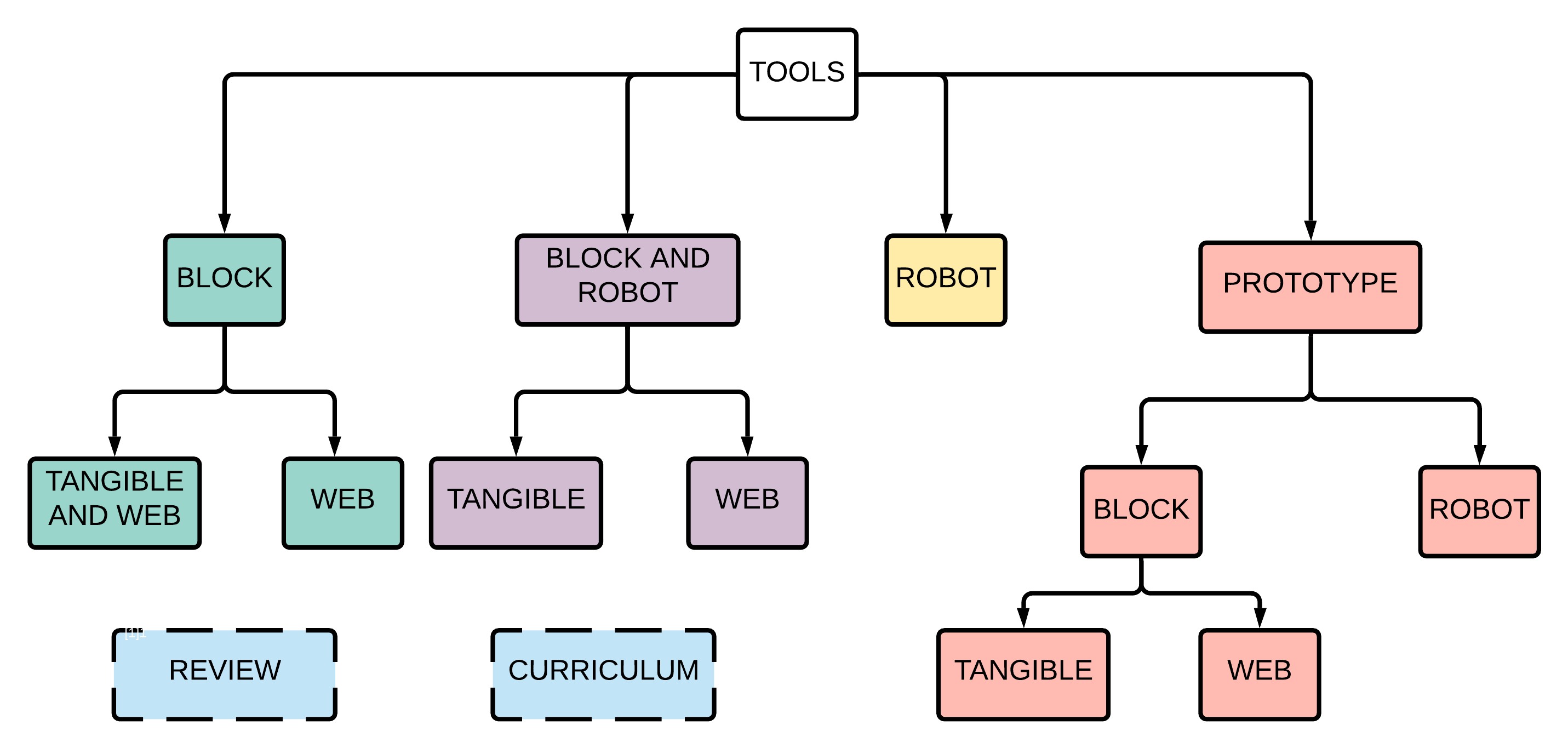}
    \caption{Proposed taxonomy.}
    \label{fig:taxonomy}
\end{figure}

In this section, we propose a taxonomy that can be seen in Figure \ref{fig:taxonomy}. There are 3 main categories, curriculum, review, and tools. The review and curriculum have already been presented, and this section emphasis on the tools. In tools, there are four possible types: block, robot, block and robot, and prototype. Whenever there is a tool based on blocks, it may be tangible or web. Despite having the taxonomy organized that way, it is possible to have papers only classified in one, another, or more than one kind of tool.


\subsubsection{Block}



\noindent\textbf{Web}

In Papadakis et al. \citep{Papadakis2016} a case study carried out in 2016 on the ScratchJr tool is presented. The study sample consisted of 43 preschool children (22 boys, 21 girls) who were attending classes in public and a private kindergarten in the region of Crete, Greece, during the school year 2014–2015. Findings reveal that ScratchJr enhances student interest by making the learning experience fun. Similarly, animated scenarios showed high levels of engagement among students. Specifically, ScratchJr allowed children to engage in deep reflection as they solved problems and collaborated with their peers, both of which activities enhanced their learning experience. Part of the authors from \citep{Papadakis2016} published another paper in the same way in \citep{Papadakis2019}. This new paper focused on the future teacher of kindergarten, to create the CT on the people will learn in the future for the kids. The authors adopted Scratch as the introductory programming language for a semester in the Department of Preschool Education in the University of Crete. The aim of using Scratch was to excite students' interest and familiarise them with the basics of programming. For 13 weeks, students were introduced to the main Scratch concepts and were asked to prepare their projects afterward. For the projects, they were required to develop a game to teach specific concepts about Mathematics or Physical Science or present an Aesop myth to preschool age students. The results we obtained were more satisfactory than expected and, in some regards, encouraging.

The authors of \citep{Lowe2019} looks at the intersection of CT and computer science in first-grade learners who are developing computational solutions involving literacy tasks. Students retell a story by animating characters in ScratchJr by breaking down the story, creating an animation storyboard, and finally implementing the plan in ScratchJr. For most of the participants, this is their first time using ScratchJr or any programming language. Therefore, their early experience with technology means they are working on an analysis of a story using literacy skills, considering a visual representation of the story, and learning how to realize the story line's expression using a computer language. Despite to did not be directly applied to the early age children, it is an excellent motivation for kindergarten teachers. 

In Ciftci et al. \citep{cciftci2020effect} the researchers evaluated an experimental study carried out to put forth the impact on the problem solving and cognitive abilities of computer programming courses applied on 4-5-year-old preschool children. This study uses a pretest-posttest control group experiment model.  According to their results, there is an increase in the non-verbal cognitive abilities of children in the experiment group with no statistically significant problem-solving skills. The `course A' from code.org was applied to the evaluation, with the help of unplugged activities (e.g., happy maps) and programming block-based.

\noindent\textbf{Tangible and Web}

In Clarke-Midura et al. \citep{ClarkeMidura2019a}, the authors examine 3 block-based coding tools applying a framework developed based on Gibson’s theory~\citep{adolph2015gibson} of affordances and Palmer’s external representations \citep{palmer1978fundamental}. They intend to verify known tools for children in early ages, motivated by more work and solutions for the development of computational thinking only for children in K-12. It is considered an interesting concept, which facilitates the child's interaction with the program, the TUI (Tangible User Interface) The tools compared are: ScratchJr, Osmo Coding Awbie\footnote{\url{https://www.playosmo.com/en/coding/}} and KIBO.

\subsubsection{Robot}

In Gonzalez et al. \citep{Gonzalez2017}, the authors show three different robotic technology applied to early school age. There are six groups of students, with a total of 131 students and eight teachers distributed in first, second, and third kindergarten. With the ease of access to groups of students and teachers, the total population will be used for the study. On the conclusions, the authors received positive results regarding the acceptance and motivation to use educational activities mediated by programmable robots in students.


Roussou et al. \citep{Roussou2020} present a research based on a case study that investigates the impact of robotics on the cultivation of CT skills in early childhood through an educational intervention implemented in a typical public kindergarten in Athens, Greece. On the material, the authors adopt the tool Code \& Go Robot Mouse Activity Set from Learning Resources\footnote{\url{https://www.learningresources.com/code-gor-robot-mouse-activity-set}}, a similar to Bee-Bot tool. The authors investigate using a pretest, intervention, and posttest, proving the benefits of robotics in kindergarten. 

\subsubsection{Block and Robot}

\noindent \textbf{Tangible}

In Urlings et al. \citep{isi2019}, sixty-five kindergarteners received assignments to go through a maze with a programmable robot, the Bee-Bot. The authors conducted this study via observation, quantifying which time and errors occurred, measuring how it was the increase of comprehension and evolution from a child with mean age equals to 6. The results were satisfactory, with a low rate of errors and a significant increase in the child's motivation.

Bers et al. \citep{Bers2019a} evaluated a ``coding as a playground'' experience in keeping with the Positive Technological Development (PTD) framework with the KIBO robotics kit, specially designed for young children. The research was conducted with preschool children aged 3–5 years old (N = 172) from three Spanish early childhood centers with different socio-economic characteristics and teachers of 16 classes. Results confirm that it is possible to start teaching this new literacy very early (at three years old). The results show that the strategies used promoted communication, collaboration, and creativity in classroom settings. Teachers had excellent experience-enhancing his/her motivation and confidence to motivate the CT on the students. The concept of ``coding as a playground'' as a new literacy, a new language for children where they can learn to code at a young age through fun, play, and creativity \citep{bers2017coding}.

\noindent\textbf{Web}

In Strawhacker et al. \citep{Strawhacker2018a}, the authors investigated the little is known about the relationship between a teacher’s unique instructional style and their students’ ability to explore and retain programming content. The study focuses on children aged 5-8 years. In this mixed-methods study, quantitative and qualitative data were collected from 6 teachers and 222 kindergartens through second-grade students at six schools across the United States from 2 months in 2014. All participants engaged in a minimum of two lessons and a maximum of seven lessons using the ScratchJr programming environment to introduce coding. Teachers reported on their classroom structure, lesson plan, teaching style, and comfort with technology. 

\noindent\textbf{Tangible and Web}

In Rial-Fern\'{a}ndez et al. \citep{RialFernandez2019}, the authors describe the case study carried out in a classroom of Early Childhood Education, with students of 5 years of age, with which basic programming concepts have been worked out. The results obtained indicate that they have been able to master the new vocabulary, assimilate the concepts, and work for themselves with the chosen tool to carry out the intervention. The tools evaluated were: Bee-Bot, Happy Maps, and another sheet with arrows that they have to cut out. All tools were applied to be a playful tool using blocks to increase the CT in programming.

In Otterborn et al. \citep{Otterborn2020a}, it was conducted systematically investigates how Swedish preschool teachers implement programming activities in their teaching practice. Data were collected through a national online survey with 199 participants. Findings revealed a range of apps and resources used in combination with tablets, where activity integration takes place as unplugged programming, digital programming, or a combination of the former. On the survey, the most tools cited by the teachers of kindergarten were: Bee-Bot, Blue-Bot, LightBotJr\footnote{\url{https://lightbot.com/}}, and ScratchJr. Another relevant result is that the study showed that nine preschool teachers use unplugged programming to introduce digital programming. 

In Kanbul et al. \citep{Kanbul2017}, the aim is to reveal the importance of coding education and robotic applications for achieving 21st-century skills in North Cyprus. The skills from pre-school level mapped by the authors to help in CT development are: (i) Skills: putting in order, separating into little pieces, giving order; (ii) Software: ScratchJr, Code.org, the first two periods of Kodable, The foos; (iii) Robotic tools: BeeBots and KIBO. In the last, the authors strongly recommend the use of tools from early ages to increase de CT.

In Pugnali et al. \citep{Pugnali2017} is investigated the impact of the interface to help and adoption of technology tools in CT. Children from 4 to 7 years old participated in the survey, evaluating ScratchJr at iPad and KIBO as the tangible tool. Results suggest that type of user interface does have an impact on children’s learning but is only one of many factors that affect positive academic and socio-emotional experiences. Tangible and graphical interfaces each have qualities that foster different types of education.


\subsubsection{Prototype}

\noindent \textbf{Robot}

In \citep{Tejera2019} the authors present a prototype solution called Robotito, which comprises the development of programming and robotics skills. The proposal has a more technical bias and uses ROS (Robot Operating System) standard mechanisms. Despite the work elucidating computational thinking as a focus, its presentation is more technical than discussing the impact of the solution on learning at an early age.

The authors of \citep{Coiro2020} propose an open-source robot platform called Pro-Robot. They aim to foster CT abilities on preschool children, including a first approach on the built platform, an embedded processing unit for not requiring any further equipment, and work on development to minimize costs. The paper is more technical but very important to increase the options of robot solutions for CT.

\noindent\textbf{Block / Tangible}

In Wang et al. \citep{Wang2014} the authors present an economic tangible programming tool called T-Maze for children aged 5-9 to build computer programs in maze games by placing wooden blocks. Through the use of computer vision technology, T-Maze provides a live programming interface with real-time graphical and voice feedback. However, as highlighted by the authors, the focus is on children aged 5-9 years, which imposes difficulties to be used by children in early ages.

\noindent\textbf{Block / Web}

Koracharkornradt in \citep{Koracharkornradt2017} presents a programming game called Tuk Tuk that aims to assist in the development of CT in children in kindergarten (a junior version of the game). It allows the child to organize the blocks to create the steps of a car until they complete a race, accumulate points, and complete a detailed task. The idea is to develop an understanding of algorithms. There is the version for older children, a rating similar to ScratchJr and Scratch.

In Kanaki et al. \citep{Kanaki2018} present the computational environment PhysGramming\footnote{\url{http://physgramming.edc.uoc.gr/programming_en.html}}, which was designed to be used by children of early childhood age, between 4 and 8 years old. PhysGramming deploys an hybrid schema of visual and text-based programming techniques, with emphasis on object-orientation, in order to introduce elementary programming concepts in early childhood education. The solution can provide three kinds of games: puzzles, matching games and group games. In each game, instructions about its functionality are given through the use of an animated cat. An example of application is the teacher creating digital games in order to help the students learn the names of some animals of the jungle, together with their nutritional habits. The authors have the ambition to present basic concepts of object orientation programming in a playful way. Despite being a recent prototype, tools in the same vein are excellent candidates for help in the development of the CT.

Barat\`{e} et al. \citep{Barate2019} show in this paper a recent evolution of a web prototype conceived initially to teach music and CT to preschool and primary school learners through a gamification approach. The software tool, Legato, is based on the metaphor of building blocks whose characteristics (e.g., position in space, shape, and color) can be associated with basic music parameters (e.g., pitch, rhythm, and timbre). Legato is a web app written using standard languages, such as HTML5, CSS, and JavaScript; besides, it adopts the Web MIDI API\footnote{\url{https://www.w3.org/TR/webmidi/}} to produce sounds. The prototype is made publicly available for evaluation and uses in an educational context. The paper did not restrict the age of the child to use it. However, it is a technical specification. The main objectives induce to be used only for kids after 5-years old, once it is necessary to manipulate the blocks to construct the music on a web-based platform.

CodyColor \citep{Klopfenstein2019} is a simplified coding game, which takes basic programming instructions representing movements (e.g. turn left and turn right) and represents them with color blocks. In contrast to most other coding games, color-coded programming relies on no symbolic interpretation on part of the player in order to be approachable by very young players as well. This initiative is based on Hour of Code CodyRoby, and it is a new version of CodyColor with massive multiplayer support.

\section{Discussion}
\label{sec:discussion}

This paper presented state of the art in CT at an early age based on a systematic review. Considering only children between 2 and 5 years old were highlight the main tools and curriculum proposals and evaluation. A taxonomy was introduced to show the most gaps and advances in the last decade (2010-2020).

Based on the 34 selected works from the Literature Systematic Review, it was needed to perform a pre-processing step, where data organization, enrichment, consolidation and  formatting are performed. Specifically, the titles of all works were fragmented in words, occurrence, and next, exclusion of stop-words (words without semantic relevance) such as conjunction and prepositions. The terms of interest were unified through a process of radical reduction or equivalence to the most frequent similar word.

\begin{table}{}
\centering
\caption{Frequency of words most used.}  
\label{wordFreq}
\begin{tabular}{p{0.5cm}llll}
\\ \hline
\textbf{Id} & \textbf{Word} & \textbf{Count} & \textbf{(\%)}  \\ \hline
1 & Computational & 15& 44.1 \\ 
2 & Thinking & 15& 44.1 \\ 
3 & Programming & 12& 35.2 \\ 
4 & Robotics & 10& 29.4 \\ 
5 & Teaching & 8& 23.5 \\ 
6 & Coding & 7& 20.5 \\ 
7 & Development & 7& 20.5 \\ 
8 & Preschool & 7& 20.5 \\ 
9 & Learning & 6& 17.6 \\ 
10 & Childhood & 6& 17.6 \\ 
11 & Teacher & 5& 14.7 \\ 
12 & Kindergarten & 5& 14.7 \\ 
13 & Early & 5& 14.7 \\ 
14 & Children & 5& 14.7 \\ 
15 & Educational & 4& 11.7 \\ 
16 & Computer & 4& 11.7 \\ 
17 & Activities & 4& 11.7 \\ 
18 & Education & 4& 11.7 \\ 
19 & With & 4& 11.7 \\ 
20 & Skills & 3& 8.8 \\ \hline
\end{tabular}
\end{table}

\begin{table}{}
\centering
\caption{Ranking of most relevant words by year.} 
\label{wordByYear}

\begin{tabular}{p{3.5cm}lrrrr}
\\ \hline
\textbf{Word ($\%$)} & \textbf{2017} & \textbf{2018} & \textbf{2019} & \textbf{2020}  \\ \hline
Computational & 50   & 33   & 35   & 25  \\ 
Thinking & 50   & 33   & 35   & 25  \\ 
Programming & 50   & 67   & 24   & 0  \\ 
Robotics & 17   & 0   & 29   & 50  \\ 
Teaching & 0   & 67   & 29   & 25  \\ 
Coding & 17   & 0   & 29   & 25  \\ 
Development & 17   & 0   & 24   & 25  \\ 
Preschool & 33   & 33   & 24   & 100  \\ 
Learning & 33   & 33   & 18   & 0  \\ 
Childhood & 0   & 0   & 29   & 0  \\ 
Teacher & 17   & 33   & 18   & 0  \\ 
Kindergarten & 0   & 0   & 0   & 0  \\ 
Early & 0   & 0   & 24   & 0  \\ 
Children & 17   & 33   & 0   & 25  \\ 
Educational & 17   & 0   & 6   & 50  \\ 
Computer & 0   & 0   & 18   & 25  \\ 
Activities & 0   & 0   & 24   & 0  \\ 
Education & 17   & 33   & 6   & 0  \\ 
With & 0   & 0   & 18   & 0  \\ 
Skills & 17   & 0   & 6   & 25  \\ \hline
\end{tabular}
\end{table}


Table \ref{wordFreq} shows the 20 most frequent words among the titles of the works. In highlight, with 15 occurrences, the words Computational and Thinking appear in about 44\% of the analyzed papers. In addition, these terms were broken down by year, starting in 2017. Due to the low amount of publications, the years 2010, 2014 and 2016 were not reported in Table \ref{wordByYear}. Still based on this table, for example, the term \text{Preschool} is present in all works published in 2020, while the terms \textit{Robotics} and \textit{Educational}, also with high occurrence, appear in half of the publication titles.


In the pre-processing stage, the consolidation of authors was performed in order to identify frequent and interesting researchers in the context of the present work. Due to typos, abbreviations and omissions of part (s) of the surname, there was a human verification step to decide, in case of doubt, if two names correspond to the same author. Table \ref{frequentAuthor} lists the authors who have more than one work among those considered in this research. It is noteworthy that, although the search did not consider specific authors, six works by researcher Marina Umaschi Bers (Bers, M.U.) were among the selection made by the systematic review, which indicates its strong performance and influence in the area.

\textcolor{black}{
It is clear that this field needs attention, and the last five years prove that a growing number of research, papers, and new initiatives have flourished. The significant amount of papers in journals led us to believe it is an excellent opportunity to motivate the participants to offer conferences and events where these new finds could be discussed more thoroughly. Another important observation was the effort conducted by many countries, such as USA and Spain to incorporate, in kindergarten's curriculum, the CT as an improvement for children's capabilities in a variety of fields in their life. Finally, the most common way to validate a proposal, tool etc., is to conduct a use case following pre-test and post-test in different cities and/or countries. 
}

\textcolor{black}{
This research's main question (i.e., ``Which tools and strategies were adopted to encourage the CT development at early ages?'') was answered. It is clear to see that there are many initiatives, however, a huge part of them is limited in terms of impact on people's life. Considering it, the authors suggest conducting more experiments and evaluations with tools and strategies for CT development at early ages. These future works need to consider the actual and long-term impacts in the life of these students.”
}

\begin{table}[t]
\centering
\caption{Most frequent authors from papers discussed.} 
\label{frequentAuthor}

\begin{tabular}{p{0.5cm}lrrr}
\\ \hline
\textbf{Id} & \textbf{Author} & \textbf{Total} & \textbf{$\%$}   \\ \hline
1 & Bers, M.U. & 7 & 21 \\ 
2 & Kalogiannakis, M. & 3 & 9\\ 
3 & Papadakis, S. & 2 & 6\\ 
4 & Sullivan, A. & 2 & 6\\ \hline
\end{tabular}
\end{table}

Furthermore, there is an ongoing discussion about gender disparity in STEM fields, where man outnumber women \citep{sullivan2019investigating}. Usually, interventions aiming to increase interest of women in science related fields are performed during high school and collage with subpar results. New evidence shows that introducing an appropriate curriculum with CT can increase girls' interest in engineering. 

Another important issue that must be discussed is how the technology will affect the future of the next generation of kids. One of the primary drivers of change in the current world is Artificial Intelligence (AI) and automation. Job characteristics and required skills will change drastically because of it. These bring us a variety of opportunities related to social and economic mobility, all that could lead us to a better society. But in order to reap all this possible benefits we must sow our future generations with this new knowledge.

The World Economic Forum \citep{World:2009} discussed how artificial intelligence is shaking up the job market and two trends were perceived: one showing the continuous need for tech jobs and skills; and one pointing to human-centric skills, which directly depends on human qualities. Also, the impact of AI is not just theoretical anymore and its influence can be observed across industries and jobs worldwide. There are way more information about this matter in different types of media, but this discussion is not the center of this work.

Thus, the aforementioned points reinforce the need for an early introduction of CT in order to prepare this future generation to a new reality regarding the job market. To conclude, the authors believe that the first step to improving the CT at an early age is to start using these resources as soon as possible.


\section*{Acknowledgement(s)}

Blind version


\printcredits

\vspace{-6pt}

\bibliographystyle{apalike}
\bibliography{cas-refs.bib}

\end{document}